\documentclass[]{IEEEtran}
    
\usepackage{amsmath}
\usepackage{graphicx} 
\usepackage{epstopdf}
\usepackage{amssymb}
\usepackage{amsfonts}
\usepackage{amsthm}
\usepackage{cite}
\usepackage{bm,comment}
\usepackage{algorithm}
\usepackage{algpseudocode}

\usepackage{subeqnarray}
\usepackage{subfigure}
\usepackage{multicol}
\usepackage{multirow}
\usepackage{diagbox}
\usepackage{slashbox}
\usepackage{stfloats}
\usepackage{float}
\usepackage{color} 
\usepackage{cases}

\usepackage{lipsum}

\newtheorem{proposition}{\bf{Proposition}}

\newtheorem{lemma}{\bf{Lemma}}

\usepackage{geometry}
\usepackage[bookmarks=true, colorlinks=true, breaklinks=true, urlcolor =black]{hyperref}
\begin{document}
\setlength{\textfloatsep}{4pt}
\title{Near-Field Wideband Secure Communications: \\
An Analog Beamfocusing Approach}

\author{Yuchen Zhang, \emph{Graduate Student Member, IEEE}, Haiyang Zhang, \emph{Member, IEEE}, \\Sa Xiao, \emph{Member, IEEE}, Wanbin Tang, \emph{Member, IEEE}, and Yonina C. Eldar, \emph{Fellow, IEEE}
\thanks{
Yuchen Zhang, Sa Xiao, and Wanbin Tang are with the National Key Laboratory of Science and Technology on Communications, University of Electronic Science and Technology of China, Chengdu, 611731 China (e-mail: yc\_zhang@std.uestc.edu.cn, {xiaosa,wbtang}@uestc.edu.cn).

Haiyang Zhang is with the School of Communication and Information
Engineering, Nanjing University of Posts and Telecommunications, Nanjing
210003, China (e-mail: haiyang.zhang@njupt.edu.cn).

Yonina C. Eldar is with the Faculty of Mathematics and Computer Science, Weizmann Institute of Science, Rehovot 7610001, Israel (e-mail: yonina.eldar@weizmann.ac.il).

}}

\maketitle

\begin{abstract}
In the rapidly advancing landscape of 6G, characterized by ultra-high-speed wideband transmission in millimeter-wave and terahertz bands, our paper addresses the pivotal task of enhancing physical layer security (PLS) within near-field wideband communications. We introduce true-time delayer (TTD)-incorporated analog beamfocusing techniques designed to address the interplay between near-field propagation and wideband beamsplit, an uncharted domain in existing literature. Our approach to maximizing secrecy rates involves formulating an optimization problem for joint power allocation and analog beamformer design, employing a two-stage process encompassing a semi-digital solution and analog approximation. This problem is efficiently solved through a combination of alternating optimization, fractional programming, and block successive upper-bound minimization techniques. Additionally, we present a low-complexity beamsplit-aware beamfocusing strategy, capitalizing on geometric insights from near-field wideband propagation, which can also serve as a robust initial value for the optimization-based approach. Numerical results substantiate the efficacy of the proposed methods, clearly demonstrating their superiority over TTD-free approaches in fortifying wideband PLS, as well as the advantageous secrecy energy efficiency achieved by leveraging low-cost analog devices.
\end{abstract}
\begin{IEEEkeywords}
Physical layer security, near-field communications, wideband communications, beamfocusing.
\end{IEEEkeywords}

\IEEEpeerreviewmaketitle
\section{Introduction}
The explosive growth of mobile networks has facilitated massive connections and extended coverage in the air, which has brought immense convenience to daily life, enabling ubiquitous communication and the seamless transmission of information everywhere. However, this remarkable advancement in wireless communication also raises a crucial concern about security. The open nature of wireless communications exposes transmitted information to potential eavesdropping by unauthorized entities.
Since the seminal work of Wyner \cite{wyner1975bell}, the concept of physical layer security (PLS) has emerged as a vital means of preventing unauthorized decoding and safeguarding transmissions. It has gained significant attraction and sparked intensive research efforts, as a complement to cryptography. Thanks to the development of multi-antenna technology, beamforming has emerged as an efficient approach to enhance PLS by leveraging spatial degrees of freedom (DoFs)\cite{physec2017survey}. The core idea revolves around forming directional beams towards legitimate users while minimizing energy leakage towards suspicious nodes through the coordination of multiple antennas. This strategy results in a substantial improvement in PLS\cite{physec2017survey}. Meanwhile, the evolution of modern communications towards 6G and beyond envisions ultra high-speed wideband transmission in the very high frequency bands, facilitated by extremely large-scale antenna array. This introduces phenomena such as near-field effect\cite{nfc2023review}, beamsquint\cite{wukai2019jstsp}, and beamsplit\cite{dai2022twc,ruochen2023tcom}. Against this background, there is a strong desire for specialized designs capable of effectively harnessing or addressing these issues, thereby boosting the PLS of 6G near-field systems.

\subsection{Previous works}
To fulfill the drastically increasing data rate demands of 6G and beyond, wireless communication systems are evolving into high-frequency millimeter-wave (mmWave) and even terahertz (THz) bands, where substantial bandwidth resources are readily available\cite{ning2023ojsoc}. However, the challenge lies in compensating for the severe propagation loss that occurs within these high-frequency channels. One potential solution involves expanding the antenna array by incorporating more elements that enables the creation of directional transmissions with higher beamforming gain. Nonetheless, the deployment of a large-scale antenna array, coupled with the use of extremely small wavelengths, significantly extends the Rayleigh distance that is proportional to the aforementioned factors\cite{nfc2023review}. Rayleigh distance is a crucial metric that distinguishes between far-field and near-field transmission. Transmission beyond the Rayleigh distance upholds conventional far-field plane wavefronts assumptions, allowing for beamforming in a specific direction. However, when transmission occurs within the Rayleigh distance, it ushers in a near-field effect, prompting a shift in transmission model towards spherical wavefronts as a more fitting representation. This transformation facilitates the creation of angle-distance-dependent beams, usually referred to  beamfocusing\cite{haiyang2022twc}, directed towards specific positions, which opens up the possibility of mitigating inter-user interference, even in scenarios where users are situated in close directions\cite{haiyang2022twc}. The introduced novel distance-wise degree of freedom (DoF) has been exploited in prior research to improve the performance of various near-field systems such as downlink multiple access\cite{zidong2023jsac}, beam training\cite{cunhua2023tcom}, wireless power transfer\cite{haiyang2022cmag}, and radio localization\cite{henk2023twc}.


The implementation of massive antenna arrays in high-frequency systems faces significant challenges due to their prohibitively high hardware costs and power consumption\cite{chong2021wcmag}. Traditional fully digital architectures, where each antenna element connects to a dedicated radio frequency (RF) chain, are impractical for such systems\cite{molisch2017cmag}. To strike a balance between performance and cost, hybrid analog and digital array architectures are preferred in mmWave/THz systems\cite{ahmed2018st,yonina2019tsp}. These typically fall into three categories: fully-connected, partially-connected, and dynamically-connected architectures\cite{longfei2020jsac}.
Fully-connected setups require complex antenna-crossing transmission lines, while dynamically-connected designs rely on antenna-switching networks, both leading to undesirable wiring and heat issues\cite{heatproblem2015}. In contrast, the partially-connected array architecture, also known as array of sub-arrays, offers an appealing low-cost solution, where the antenna elements are grouped, sharing a common RF chain\cite{goeffrey2016cmag}. Within each sub-array, individual phase shifters (PSs) enable analog beamforming by jointly adjusting their phase shifts\cite{ning2023ojsoc,longfei2020jsac,goeffrey2016cmag}. 


In the realm of ultra high-speed communications with extensive signal bandwidth, frequency-dependent precoding, realized by fully-digital architecture, is favourable to be adaptive to wideband frequency-selective channels. However, the practical implementation of massive antenna arrays can introduce significant beamsplit due to the analog front-end in hybrid fabrication, which may result in significant performance degradation. For instance, in far-field transmission, a beamformer designed for a specific direction at a given frequency might experience substantial deviation to an unintended direction due to frequency mismatches\cite{dai2022twc}, thus compromising the beamforming gain over the entire bandwidth. To mitigate the effect of beamsplit, recent studies\cite{dai2022twc,ruochen2023tcom,zhai2021jsac,feifei2023tcom} have employed true-time delayers (TTDs), which introduce specific time delays to signals, thereby creating frequency-dependent phase shifts. These hybrid beamforming architectures, incorporating a limited number of TTDs alongside a larger array of PSs, have demonstrated significant advantages in enhancing far-field wideband communications performance\cite{dai2022twc,ruochen2023tcom,zhai2021jsac,feifei2023tcom}. This improvement holds true even when employing cost-effective hardware options like fixed-time-delay TTDs\cite{longfei2022jsac}. More recently, there have been several attempts extending the study of beamsplit to near-field communications\cite{heath2022twc,wang2023arxiv,cui2023arxiv,yz2023wcl}, exploring spatial waveform design\cite{heath2022twc}, heuristic piece-wise approximation\cite{cui2023arxiv}, convex optimization\cite{wang2023arxiv}, and deep learning\cite{yz2023wcl}, respectively.

\subsection{Motivation and Contributions}
Given the evolving landscape towards 6G, there is a scarcity of research addressing aforementioned challenges and opportunities within PLS. In response to the demand for cost-effective solutions in practical systems, recent efforts have focused on secure analog and hybrid beamforming techniques\cite{analogbf2017,analogbf2017tvt,analogbfspl2017,analogboyd2020}, facilitated by economical devices like PSs. However, these efforts primarily focused on far-field narrowband transmission.
The authors of \cite{pls2023arxiv} initially explored PLS in the near-field region, highlighting the potential to enhance security even when the eavesdropper is in close angular proximity to legitimate receiver. Nevertheless, their investigation was still confined to narrowband transmission.
In \cite{wbpls2016globecom} and \cite{wbpls2017tcom}, the focus was on realizing wideband secure mmWave communications through the use of hybrid beamforming. However, these studies lacked specific designs for harnessing the near-field effect and mitigating beamsplit. 

To bridge the gap, this paper explores near-field wideband secure communications and presents efficient analog beamfocusing approaches to enhance PLS. The main contributions are summarized as follows. 
\begin{itemize}
\item To the best of the authors' knowledge, this is the first work to investigate the interplay between near-field propagation and wideband beamsplit concerning PLS. By utilizing an analog structure comprising TTDs and PSs, we propose efficient beamfocusing approaches that exploit near-field propagation while mitigating beamsplit, ultimately enhancing PLS performance.
\item To maximize secrecy rate, we formulate an optimization problem that jointly allocates power and designs the analog beamformer, which is solved in a two-stage manner. First, we apply an alternating optimization (AO) framework to tackle the non-convex semi-digital counterpart of the original problem, optimizing the beamformer iteratively through fractional programming (FP). In the second stage, the analog beamformer is obtained by configuring the time delays of TTD and the phases of PS alternatingly to mimic the semi-digital beamformer. Here, we employ the block successive upper-bound minimization (BSUM) framework to circumvent the challenge of tuning time delays.
\item We propose a low-complexity beamsplit-aware beamfocusing approach, which capitalizes on the geometric information inherent in near-field wideband propagation. 
TTDs and PSs are heuristically configured, guided by the beamsplit trace equation. Furthermore, the obtained values can serve as initial values for the optimization-based approach, strengthening its robustness.  
\item Numerical results confirm the effectiveness of our proposed analog beamfocusing strategies, showcasing their superiority over TTD-free approaches. These strategies significantly enhance PLS in near-field wideband transmission, emphasizing the achieved energy efficiency through the use of low-cost analog devices.
\end{itemize}

\subsection{Organization and Notations}
The remaining of the paper is organized as follows. The system model is described in Section II. In Sections III and IV, we present the optimization-based and low-complexity geometry-based analog beamfocusing approaches, respectively. Numerical results are presented in Section V. Finally, Section VI concludes the paper.

The main notations throughout this paper are clarified as follows. Regular lowercase letters denote scalars, bold lowercase letters represent vectors, and bold uppercase letters signify matrices, with $|a|$ denoting the absolute value of scalar $a$ and $\left\|\mathbf{a}\right\|$ representing the 2-norm of vector $\mathbf{a}$. Additionally, $\lfloor a \rfloor$ and $\lceil a \rceil$ are the rounding operators that round $a$ to the nearest integer towards $-\infty$ and $+\infty$, respectively. Superscripts $T$ and $H$ indicate the transpose and Hermitian transpose of a vector or matrix. Furthermore, $\mathrm{tr}(\mathbf{A})$ and $\mathrm{rank}(\mathbf{A})$ stand for the trace and rank of matrix $\mathbf{A}$, while $\mathbf{A}\succeq \mathbf{0}$ implies that matrix $\mathbf{A}$ is Hermitian and positive semi-definite. The real part of a scalar $a$ is denoted as $\mathrm{Re}[a]$, and $[a]^{+}$ is defined as $\max(0, a)$. $\text{diag}(\mathbf{a})$ represents a diagonal matrix where the diagonal elements are formed by the elements of $\mathbf{a}$ while $\text{card}(\mathcal{M})$ denotes the cardinality of the set $\mathcal{M}$. Additionally, $\mathcal{CN}(\boldsymbol{\mu},\mathbf{C})$ represents the circularly symmetric complex Gaussian (CSCG) distribution with mean $\boldsymbol{\mu}$ and covariance matrix $\mathbf{C}$.

\begin{figure}[!t]
\centering
\includegraphics[width=0.48\textwidth]{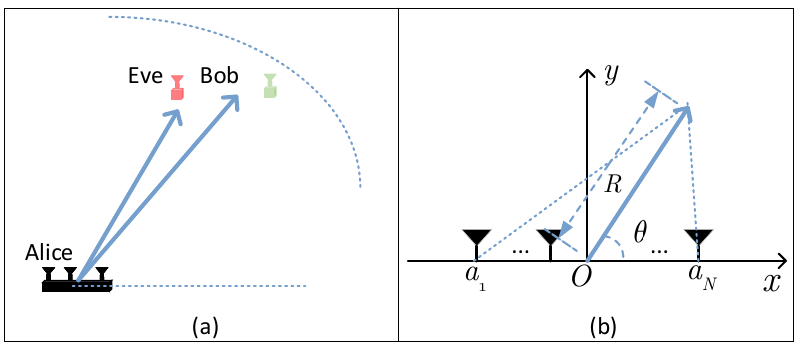}
\caption{(a) System model. (b) Near-field transmission. }
\label{sys-mod-I}
\end{figure} 

\section{System Model}
As shown in Fig. \ref{sys-mod-I} (a), we consider a multi-input single-output single-eavesdropper wiretap channel, where Alice equipped with an uniform linear array (ULA) of $N$ elements transmits confidential messages to single-antenna Bob under the eavesdropping of the single-antenna Eve. We assume that both Bob and Eve are located in the near-field region of Alice. Instead of a conventional narrowband assumption, we take the demand of wideband transmission into account. In the following, we introduce the near-field channel model, sketch the wideband transmission model, and formulate the optimization problem to maximize the secrecy rate of the system, respectively. 
 
\subsection{Near-Field Channel Model}
In contrast to far-field transmission where only angle-related aspects matter, the near-field region introduces variations in both angle and distance for each antenna's propagation path to the destination. Consequently, the channel is characterized by modeling the geometric relationships within the Cartesian coordinate system.
As illustrated in Fig. \ref{sys-mod-I} (b), the ULA is parallel to the $x$-axis with its center  located at the origin point $O$. 
Let $a_n$ be the $x$-coordinate of the $n$-th antenna. According to geometric knowledge, the distance between the $n$-th antenna and a node located at $(R,\theta)$ in the polar coordinate system can be calculated as 
\begin{equation}\label{distance_n}
D_n = \sqrt{a_n^2+R^2-2a_nR\cos\theta}.    
\end{equation}
The corresponding time-domain channel impulse response accounting for the path loss is formulated as
\begin{equation}\label{time_impulse_n}
h_n\left(t,R,\theta\right) = \frac{c}{4\pi f D_n}\delta\left(t-\frac{D_n}{c}\right),    
\end{equation}
where $c$ and $f$ are the speed of light and frequency of the transmitted signal, respectively. By applying the Fourier transform to \eqref{time_impulse_n}, the frequency-domain channel is given by
\begin{equation}\label{freq_impulse_n}
h_n\left(f,R,\theta\right) = \frac{c}{4\pi f D_n}e^{\frac{-j2\pi f D_n}{c}}.
\end{equation}
As a result, the frequency-domain channel vector in regard to position $(R,\theta)$ is expressed as
\begin{equation}\label{freq_impulse_vec}
\mathbf{h}\left(f,R,\theta\right) = \left[h_1\left(f,R,\theta\right),\ldots,h_N\left(f,R,\theta\right)\right]^T.
\end{equation}

\begin{figure}[!t]
\centering
\includegraphics[width=0.48\textwidth]{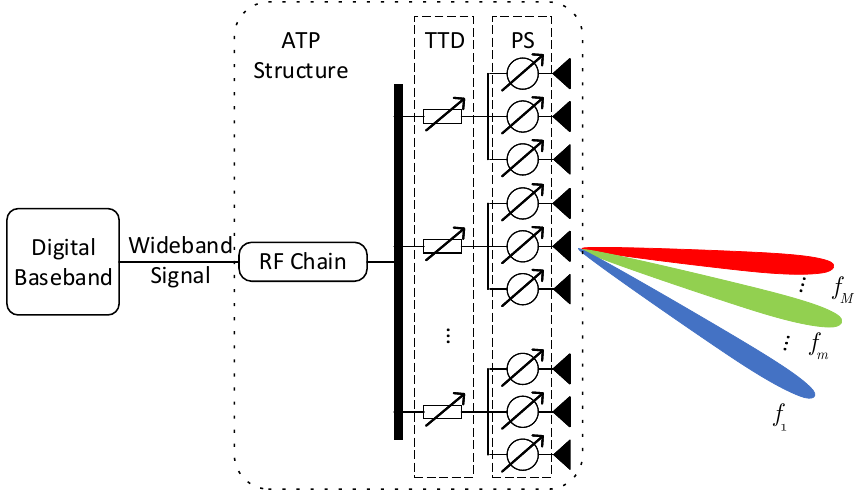}
\caption{Illustration of ATP structure for wideband transmission. }
\label{sys-mod-II}
\end{figure} 
 
\subsection{Wideband Transmission Model}
We consider orthogonal frequency division multiplexing (OFDM)-based wideband transmission. The $m$-th baseband subcarrier frequency is given by
\begin{equation}\label{ofdm_freq}
\widetilde{f}_m=-\frac{B}{2}+\frac{\left(m-1\right)B}{M-1},m\in \mathcal{M}\buildrel \Delta \over=\left\{1,2,\ldots,M\right\},    
\end{equation}
where $B$ and $M$ denote the bandwidth of the signal and the number of subcarriers, respectively. The baseband signal is first up-converted by an RF chain to carrier frequency $f_m = f_c + \widetilde{f}_m$ and then processed by an analog beamformer, where $f_c$ is the central carrier frequency. Conventional analog beamformers primarily utilize a set of frequency-independent phase shifters (PSs). However, this approach can lead to a beamsplit issue, thereby compromising the performance of wideband transmission. This is because the analog beamformer, designed for a specific direction at a particular frequency, may deviate significantly in other parts of the bandwidth\cite{dai2022twc}.


To alleviate this issue, as depicted in Fig. \ref{sys-mod-II}, we place $N_T$ TTDs in front of the PSs. Each TTD is connected to a group of $N_G = N/N_T$ PSs, thereby forming the analog TTD-PS (ATP) structure. 
For notational convenience, let $\mathbf{h}_{B,m}$ and $\mathbf{h}_{E,m}$ denote $\mathbf{h}\left(f_m,R_B,\theta_B\right)$ and $\mathbf{h}\left(f_m,R_E,\theta_E\right)$, respectively.
Let $s_m \sim \mathcal{CN}(0,1)$ and $P_m$ denote the normalized information symbol and  transmit power at the $m$-th carrier frequency, respectively.  
The signals received at Bob and Eve at the $m$-th carrier frequency are given by
\begin{equation}\label{receive_sig_B}
y_{B,m}=\sqrt{\frac{P_m}{N}}\mathbf{h}_{B,m}^{H}\text{diag}\left(\mathbf{w}\right)\mathbf{t}_m s_m + n_{B,m}    
\end{equation}
and 
\begin{equation}\label{receive_sig_E}  
y_{E,m}=\sqrt{\frac{P_m}{N}}\mathbf{h}_{E,m}^{H}\text{diag}\left(\mathbf{w}\right)\mathbf{t}_m s_m + n_{E,m},   
\end{equation}
where $\mathbf{w} \in \mathbb{C}^{N \times 1}$ is the PS-based beamformer and
\begin{eqnarray}\label{ttd_bf}
\begin{aligned}
\mathbf{t}_m = &\left[\underbrace{e^{-j2\pi f_m \tau_1},\ldots,e^{-j2\pi f_m \tau_1}}_{\text{First group, }N_G\text{ elements}},\ldots,\right.\\
&\;\;\;\;\;\;\;\left. \underbrace{e^{-j2\pi f_m \tau_{N_T}},\ldots,e^{-j2\pi f_m \tau_{N_T}}}_{N_T\text{-th group, }N_G\text{ elements}}\right]^T \in \mathbb{C}^{N \times 1}   
\end{aligned}	
\end{eqnarray}
is the TTD-based beamformer in regard to carrier frequency $f_m$ and delays $\tau_1,\ldots,\tau_{N_T}$. In addition, $n_{B,m} \sim \mathcal{CN}(0,{\sigma}^2)$ and $n_{W,m} \sim \mathcal{CN}(0,{\sigma}^2)$ denote additive white Guassian noises (AWGNs) with power ${\sigma}^2$ at Bob and Eve, respectively.

The introduction of TTDs make the cascaded analog beamformer frequency-dependent, thus enabling more flexible design of wideband transmission. When the delay budget for TTDs becomes zero, the ATP beamformer degenerates to the conventional PS-only beamformer. 

\subsection{Problem Formulation}
We assume that Alice possesses complete knowledge of the channel state information of both Bob and Eve. This assumption is realistic in scenarios where Eve actively engages in communications and collaboration with Alice\cite{physec2017survey,wbpls2016globecom,wbpls2017tcom}. By employing OFDM transmission, the communications channels at different subcarriers can be regarded as independent parallel channels whose achievable secrecy rate is given by \cite{bloch2008tit,eduard2008globecom}
\begin{equation}\label{whole_sr}
R_S = \sum_{m \in \mathcal{M}}\left[R_{B,m}-R_{E,m}\right]^{+},    
\end{equation}
where 
\begin{equation}\label{rate_B}
R_{B,m}=\log_2\left(1+\frac{P_m\left|\mathbf{h}_{B,m}^{H}\text{diag}\left(\mathbf{w}\right)\mathbf{t}_m\right|^2}{N{\sigma}^2}\right)
\end{equation}
and 
\begin{equation}\label{rate_E}  
R_{E,m}=\log_2\left(1+\frac{P_m\left|\mathbf{h}_{E,m}^{H}\text{diag}\left(\mathbf{w}\right)\mathbf{t}_m\right|^2}{N{\sigma}^2}\right).
\end{equation}

We aim to maximize the achievable secrecy rate by jointly optimizing the power allocation across subcarriers and designing the phase settings of PSs and time delays of TTDs. The optimization problem is formulated as 
\begin{subequations}\label{ori_prob}
\begin{align}
\mathop {\max }\limits_{P_m, \mathbf{w}, \tau_i} \;\;\; &R_S\label{ori_prob-obj}\\
{\rm{s.t.}}\;\;\;&\big|\left[\mathbf{w}\right]_{n}\big|=1,\forall n \in \mathcal{N}\buildrel \Delta \over=\left\{1,2,\ldots,N\right\},\label{ori_prob-ps}\\
& 0 \le \tau_i \le \chi,\forall i \in \mathcal{T}\buildrel \Delta \over=\left\{1,2,\ldots,N_T\right\},\label{ori_prob-ttd}\\
& \sum_{m \in \mathcal{M}} P_m \le P, \label{ori_prob-power}
\end{align}
\end{subequations}
where \eqref{ori_prob-ps} represents the unit-modulus requirement for the PSs, and $\chi$ and $P$ are the delay budget for each TTD and power budget for the wideband transmission, respectively. 

The presented problem is challenging to solve due to several key factors. First, the allocated power and the analog beamformers are coupled in the highly intractable objective function which is the sum of non-convex secrecy rates across subcarriers. Second, the cascaded structure of analog beamformer make it infeasible to configure PSs and TTDs independently. Third, the analog nature imposes unit-modulus constraint on $\mathbf{w}$, which is also non-convex. Furthermore, the configuration of TTDs are realized by optimizing time delays $\tau_i$, which in turn introduces frequency-specific phase components and further complicates the optimization across various subcarriers. In the following, we will develop efficient schemes to circumvent these difficulties.   

\section{Proposed analog beamformer design}
In this section, we propose a beamfocusing scheme utilizing the ATP structure, tailored for near-field wideband secure communications. Inspired by the conventional hybrid beamforming approach as in \cite{haiyang2022twc,wang2023arxiv,yu2016jstsp}, our scheme employs a two-step framework for analog beamformer design. We first obtain the semi-digital\footnote{In the semi-digital approach, beamformers remain frequency-independent and adhere to element-wise unit-modulus constraint as \eqref{digit_prob-bf}, distinct from fully-digital beamformers that solely require vector-wise energy constraint. This semi-digital configuration streamlines subsequent analog approximations in the two-step framework by minimizing beamformer mismatch. This is crucial because the ATP-based analog beamformers, while frequency-dependent, still maintain unit-modulus constraint.} beamformer by addressing its semi-digital counterpart, followed by an approximation to construct the analog counterpart. The fundamental principle underlying this approach is to minimize the gap between the analog beamformer and the optimized semi-digital one, thereby maximizing overall achievable secrecy rate in near-field wideband systems.

\subsection{Semi-Digital Solution}
We define the digitized secrecy rate as $\widetilde{R}_S = \sum_{m \in \mathcal{M}}[\widetilde{R}_{B,m}-\widetilde{R}_{E,m}]^{+}$, where $\widetilde{R}_{B,m}$ and $\widetilde{R}_{E,m}$ are obtained via substituting $\text{diag}\left(\mathbf{w}\right)\mathbf{t}_m$ in \eqref{rate_B} and \eqref{rate_E} by $\mathbf{v}_m \in \mathbb{C}^{N\times 1}$. The semi-digital counterpart of \eqref{ori_prob} with respect to $P_m$ and $\mathbf{v}_m$ is formulated as
\begin{subequations}\label{digit_prob}
\begin{align}
\mathop {\max }\limits_{P_m, \mathbf{v}_m} \;\;\; &\widetilde{R}_S =\sum_{m \in \mathcal{M}} \left[\log_2\left(\frac{N{\sigma}^2+P_m\left|\mathbf{h}_{B,m}^{H}\mathbf{v}_m\right|^2}{N{\sigma}^2+P_m\left|\mathbf{h}_{E,m}^{H}\mathbf{v}_m\right|^2}\right)\right]^{+}\label{digit_prob-obj}\\
{\rm{s.t.}}\;\;\;
&\big|\left[\mathbf{v}_m\right]_{n}\big| = 1,\forall m \in \mathcal{M},n \in \mathcal{N}, \label{digit_prob-bf}\\
& \eqref{ori_prob-power},
\end{align}
\end{subequations}
where \eqref{digit_prob-bf} is added to be coherent with the energy constraint for beamformers. Problem \eqref{digit_prob} is still hard to solve owing to the coupled $P_m$ and $\mathbf{v}_m$ in the non-convex objective function. Utilizing the AO framework, we proceed to tackle \eqref{digit_prob} via iteratively solving the subproblems of power allocation and beamformer design.  

\newcounter{MYtempeqncn}
\begin{figure*}[!b]
\hrulefill
\normalsize
\setcounter{MYtempeqncn}{\value{equation}}
\setcounter{equation}{15}
\begin{equation}\label{opt_pow}
\widetilde{P}_m^* = \left[-\frac{1}{2}\left(\frac{N{\sigma}^2}{\beta_{B,m}}+\frac{N{\sigma}^2}{\beta_{E,m}}\right)+\frac{1}{2}\sqrt{\left(\frac{N{\sigma}^2}{\beta_{B,m}}-\frac{N{\sigma}^2}{\beta_{E,m}}\right)^2 +\frac{4}{\mu}\left(\frac{N{\sigma}^2}{\beta_{B,m}}-\frac{N{\sigma}^2}{\beta_{E,m}}\right) }\right]^{+} 
\end{equation}
\setcounter{equation}{\value{MYtempeqncn}}
\end{figure*}

\emph{1) Subproblem of Power Allocation:} 
Given fixed beamformer $\mathbf{v}_m$, we define $\beta_{B,m} = |\mathbf{h}_{B,m}^{H}\mathbf{v}_m|^2$ and $\beta_{E,m} = |\mathbf{h}_{E,m}^{H}\mathbf{v}_m|^2$, respectively, for ease of presentation. In addition, note that $[\widetilde{R}_{B,m}-\widetilde{R}_{E,m}]^{+}=0$ when $\beta_{B,m} \le \beta_{E,m}$. We define $\mathcal{M}^{+}\buildrel \Delta \over=\{m|\beta_{B,m} > \beta_{E,m}, m \in \mathcal{M}\}$ and $\mathcal{M}^{-}\buildrel \Delta \over=\{m|\beta_{B,m} \le \beta_{E,m}, m \in \mathcal{M}\}$, respectively. Specifically, we have $\mathcal{M}^{+}\cup\mathcal{M}^{-}=\mathcal{M}$ and $\mathcal{M}^{+}\cap\mathcal{M}^{-}=\emptyset$. The subproblem of power allocation is transformed into
\begin{subequations}\label{digit_prob_pow}
\begin{align}
\mathop {\max }\limits_{P_m} \;\;\; & \sum_{m \in \mathcal{M}^{+}} \log_2\left(\frac{N{\sigma}^2+P_m \beta_{B,m}}{N{\sigma}^2+P_m \beta_{E,m}}\right)\label{digit_prob_pow-obj}\\
{\rm{s.t.}}\;\;\;
&\eqref{ori_prob-power}.
\end{align}
\end{subequations}
This problem is equivalent to the secure power allocation problem for single-input single-output single-eavesdropper parallel independent channel, which can be solved in closed form \cite{eduard2008globecom}. The optimal solution can be obtained via examining the Karush-Kuhn-Tucker conditions of \eqref{digit_prob_pow}, and is derived as 
\begin{eqnarray}\label{update_POWER}
\begin{aligned}
P_m^*=\left\{\begin{array}{ll}
\widetilde{P}_m^*,& m \in \mathcal{M}^{+},\\
0,&m \in \mathcal{M}^{-},
\end{array}\right.
\end{aligned}
\end{eqnarray}
where $\widetilde{P}_m^*$ is given by \eqref{opt_pow} at the bottom of this page and $\mu$ is a constant such that \eqref{ori_prob-power} holds. The insight here is to allocate power exclusively to subcarriers with a greater equivalent channel gain towards Bob, as opposed to Eve. Note that \eqref{opt_pow} is non-decreasing with respect to $\mu$, we could find $\mu$ through bisection search similar to classical water-filling scheme \cite{wf2014wcl}, which are omitted for brevity.

\emph{2) Subproblem of Beamformer Design:} 
Once $P_m$ is fixed, by defining matrices $\mathbf{H}_{B,m}=\mathbf{h}_{B,m}\mathbf{h}_{B,m}^{H}\in \mathbb{C}^{N\times N}$, $\mathbf{H}_{E,m}=\mathbf{h}_{E,m}\mathbf{h}_{E,m}^{H}\in \mathbb{C}^{N\times N}$, and $\mathbf{V}_{m}=\mathbf{v}_{m}\mathbf{v}_{m}^{H}\in \mathbb{C}^{N\times N}$, the subproblem of beamformer design is lifted as
\addtocounter{equation}{1}
\begin{subequations}\label{digit_prob_bf}
\begin{align}
\mathop {\max }\limits_{\mathbf{V}_m} \;\;\; &\sum_{m \in \mathcal{M}^{+}} \log_2\left(\frac{N{\sigma}^2+P_m\text{tr}\left(\mathbf{H}_{B,m}^{H}\mathbf{V}_m\right)}{N{\sigma}^2+P_m\text{tr}\left(\mathbf{H}_{E,m}^{H}\mathbf{V}_m\right)}\right)\label{digit_prob_bf-obj}\\
{\rm{s.t.}}\;\;\;
&\left[\mathbf{V}_m\right]_{n,n} = 1,\forall n \in \mathcal{N}, \label{digit_prob_bf-power}\\
&\mathbf{V}_m \succeq \mathbf{0},\label{ori_prob_bf-psd}\\
&\text{rank}\left(\mathbf{V}_m\right) = 1,\forall m \in \mathcal{M}^{+}\label{ori_prob_bf-rank}.
\end{align}
\end{subequations}
Subsequently, we introduce auxiliary variables $\xi_m,\forall m \in  \mathcal{M}^{+}$ to surrogate the complicated logarithmic components in the objective function. Problem \eqref{digit_prob_bf} is converted into 
\begin{subequations}\label{digit_prob_bf_convert_1}
\begin{align}
\mathop {\max }\limits_{\mathbf{V}_m,\xi_m} \;\;\; &\sum_{m \in \mathcal{M}^{+}} \xi_m \\
{\rm{s.t.}}\;\;\;
& 2^{\xi_m} \le \frac{N{\sigma}^2+P_m\text{tr}\left(\mathbf{H}_{B,m}^{H}\mathbf{V}_m\right)}{N{\sigma}^2+P_m\text{tr}\left(\mathbf{H}_{E,m}^{H}\mathbf{V}_m\right)},  \label{digit_prob_bf_convert_1-rate}\\
&\eqref{digit_prob_bf-power},\eqref{ori_prob_bf-psd},\eqref{ori_prob_bf-rank}.
\end{align}
\end{subequations}

To overcome the difficulty caused by the non-convex fractional constraint \eqref{digit_prob_bf_convert_1-rate}, we exploit the equivalence established in \cite{kaiming2018tsp} to transform \eqref{digit_prob_bf_convert_1-rate} into 
\begin{eqnarray}\label{fr-trans}
\begin{aligned}
2^{\xi_m} \le  \mathop {\max }\limits_{\lambda_m}   2\lambda_m  &\sqrt{N{\sigma}^2+P_m\text{tr}\left(\mathbf{H}_{B,m}^{H}\mathbf{V}_m\right)}\\
&-\lambda_m^2\left(N{\sigma}^2+P_m\text{tr}\left(\mathbf{H}_{E,m}^{H}\mathbf{V}_m\right)\right),    
\end{aligned}
\end{eqnarray}
where $\lambda_m$ is the introduced auxiliary variable. As constraint \eqref{fr-trans} possesses the less-than-max structure, $\lambda_m$ could be up-lifted to the objective function, thereby leading to 
\begin{subequations}\label{digit_prob_bf_convert_2}
\begin{align}
\mathop {\max }\limits_{\substack{\mathbf{V}_m,\lambda_m,\\ \xi_m}} \;\;\; &\sum_{m \in \mathcal{M}^{+}} \xi_m \\
{\rm{s.t.}}\;\;\;
& 2^{\xi_m} \le   2\lambda_m  \sqrt{N{\sigma}^2+P_m\text{tr}\left(\mathbf{H}_{B,m}^{H}\mathbf{V}_m\right)} \nonumber\\
& \;\;\;\;\;\;\;\;\;-\lambda_m^2\left(N{\sigma}^2+P_m\text{tr}\left(\mathbf{H}_{E,m}^{H}\mathbf{V}_m\right)\right), \label{digit_prob_bf_convert_2-rate}\\
&\eqref{digit_prob_bf-power},\eqref{ori_prob_bf-psd},\eqref{ori_prob_bf-rank}.
\end{align}
\end{subequations}
Note that when $\mathbf{V}_m$ is fixed, the optimal $\lambda_m$ is
\begin{equation}\label{lambda_update}
\lambda_m^* = \frac{\sqrt{N{\sigma}^2+P_m\text{tr}\left(\mathbf{H}_{B,m}^{H}\mathbf{V}_m\right)}}{N{\sigma}^2+P_m\text{tr}\left(\mathbf{H}_{E,m}^{H}\mathbf{V}_m\right)}.    
\end{equation}
On the other hand, to deal with constraint \eqref{ori_prob_bf-rank} when $\lambda_m$ is given, we present the following lemma \cite{huang2023tsp}, which can be used to lift the rank-one constraint into a more tractable matrix form. 
\begin{lemma}
For a positive semi-definite Hermitian matrix $\mathbf{V} \in \mathbb{C}^{N \times N}$, the constraint $\text{rank}\left(\mathbf{V}\right)=1$ is equivalent to the following conditions
\begin{subequations}\label{lemma2}
\begin{align}
&\text{tr}\left(\mathbf{V}\mathbf{A}\right)-2\varpi-	\text{tr}\left(\mathbf{B}\right) \ge 0,\\
&\text{tr}\left(\mathbf{A}\right)=1,\\
&\mathbf{B} - \mathbf{V} + \varpi \mathbf{I}\succeq \mathbf{0}, \\
&\mathbf{A} \succeq \mathbf{0},\\
&\mathbf{B} \succeq \mathbf{0},\label{lemma2-b}
\end{align}	
\end{subequations}
where $\varpi$ and the Hermitian matrices $\mathbf{A},\mathbf{B} \in \mathbb{C}^{N \times N}$ are the introduced auxiliary variables.
\end{lemma}

Based on Lemma 1, \eqref{digit_prob_bf_convert_2} with given $\lambda_m$ can be recast as
\begin{subequations}\label{digit_prob_bf_convert_3}
\begin{align}
\mathop {\max }\limits_{\substack{\mathbf{V}_m,\mathbf{A}_m,\mathbf{B}_m,\\
\xi_m,\varpi_m}} \;\;\; &\sum_{m \in \mathcal{M}^{+}} \xi_m \\
{\rm{s.t.}}\;\;\;
&\text{tr}\left(\mathbf{V}_{m}\mathbf{A}_{m}\right)-2 \varpi_{m}-	\text{tr}\left(\mathbf{B}_{m}\right) \ge 0,\label{digit_prob_bf_convert_3-A}\\
&\text{tr}\left(\mathbf{A}_{m}\right)=1,\label{digit_prob_bf_convert_3-B}\\
&\mathbf{B}_{m} - \mathbf{V}_{m} +  \varpi_{m} \mathbf{I}\succeq \mathbf{0},\label{digit_prob_bf_convert_3-C}\\
&\mathbf{A}_{m} \succeq \mathbf{0},\label{digit_prob_bf_convert_3-D}\\
&\mathbf{B}_{m} \succeq \mathbf{0},\forall m \in \mathcal{M}^{+},\label{digit_prob_bf_convert_3-E}\\
&\eqref{digit_prob_bf-power},\eqref{ori_prob_bf-psd},\eqref{ori_prob_bf-rank},\eqref{digit_prob_bf_convert_2-rate},
\end{align}
\end{subequations}
where $\varpi_{m}$ and $\mathbf{A}_{m}$, $\mathbf{B}_{m} \in \mathbb{C}^{N  \times N}$ are the introduced auxiliary variables.
Note that the challenge in \eqref{digit_prob_bf_convert_3} is the bilinear term of $\mathbf{V}_{m}$ and $\mathbf{A}_{m}$ in constraint \eqref{digit_prob_bf_convert_3-A}, which is non-convex. We can overcome this difficulty in an iterative manner. By fixing $\mathbf{V}_{m}$ in \eqref{digit_prob_bf_convert_3-A} to the value of the last loop, the above problem becomes convex and can be solved to update $\mathbf{V}_{m}$. Then, we carry on this process until the value of $\mathbf{V}_{m}$ converges, i.e., the mean-square error of $\mathbf{V}_{m}$ in two consecutive iterations is less than a given tolerance $\epsilon$. Specifically, given $\lambda_m$, the initial value of $\mathbf{V}_{m}$ is determined by the solution of \eqref{digit_prob_bf_convert_2} without rank-one constraint \eqref{ori_prob_bf-rank}. The convergence of this procedure is ensured, as demonstrated in \cite{huang2023tsp}. Then the optimized $\mathbf{V}_{m}$ is guarantee to be rank-one, which is used to update $\lambda_m$. Repeating this procedure until convergence, i.e., the absolute value of the difference of $\lambda_m$ in consecutive iterations is less than a given tolerance $\delta$, the solution of \eqref{digit_prob_bf} can be obtained via applying eigenvalue decomposition to the final value of $\mathbf{V}_{m}$. This completes the optimization of beamformer. The steps of solving the subproblem of beamformer design are detailed in Algorithm \ref{digt_algo_bf}.

In summary, the steps of solving \eqref{digit_prob} are presented in Algorithm \ref{digt_algo}, where $\kappa$ is the iterative tolerance of AO. 
The convergence of Algorithm 2 is guaranteed due to the monotonically increasing and bounded nature of $\widetilde{R}_S$. In each iteration of Algorithm 2, the computational complexity is dominated by Algorithm 1, i.e., the optimization of the beamformer. This complexity can be expressed as $\mathcal{O}(L_{o}((MN^2+M)^{3.5}+L_{i}(3MN^2+2M)^{3.5}))$, where $L_{o}$ and $L_{i}$ are the numbers of outer and inner iterations, respectively.

\renewcommand{\algorithmicrequire}{\textbf{Input:}}
\renewcommand{\algorithmicensure}{\textbf{Output:}}
\begin{algorithm}[t]
\caption{Beamformer Design Given Power Allocation}
\label{digt_algo_bf}       %
\begin{algorithmic}[1]
\State \textbf{Input}: $P_m$, $\delta$, $\epsilon$;
\State \textbf{Initialize}: $i=0$, $\lambda_m^{\left(0\right)}$;
\While {$\mathop {\max }\limits_{m}\|\lambda_m^{\left(i+1\right)}-\lambda_m^{\left(i\right)}\| > \delta$}
\State {$j=0$, obtain $\mathbf{V}_{m}^{\left(i,0\right)}$ via solving \eqref{digit_prob_bf_convert_2} without rank-one constraint \eqref{ori_prob_bf-rank};}
\While {$\mathop {\max }\limits_{m}\|\mathbf{V}_{m}^{\left(i,j+1\right)}-\mathbf{V}_{m}^{\left(i,j\right)}\| > \epsilon$}
\State {Obtain $\mathbf{V}_{m}^{\left(i,j+1\right)}$ via solving \eqref{digit_prob_bf_convert_3} with $\mathbf{V}_{m}$ in \eqref{digit_prob_bf_convert_3-A} setting as $\mathbf{V}_{m}^{\left(i,j\right)}$; $j = j+1$;}
\EndWhile
\State {Update $\lambda_m^{\left(i+1\right)}$ via \eqref{lambda_update}; $i = i+1$;}
\EndWhile
\State \textbf{Output}: {$\mathbf{v}_{m}$, i.e., the eigenvector with respect to the maximum eigenvalue of $\mathbf{V}_{m}$.} 
\end{algorithmic}
\end{algorithm}

\renewcommand{\algorithmicrequire}{\textbf{Input:}}
\renewcommand{\algorithmicensure}{\textbf{Output:}}
\begin{algorithm}[t]
\caption{Semi-Digital Solution via AO}
\label{digt_algo}       %
\begin{algorithmic}[1]
\State \textbf{Input}: $\kappa$, $\delta$, $\epsilon$;
\State \textbf{Initialize}: $t=0$, $\mathbf{v}_{m}^{\left(0\right)}$, $\widetilde{R}_S^{\left(0\right)}$;
\While {$\|\widetilde{R}_S^{\left(t+1\right)}-\widetilde{R}_S^{\left(t\right)}\|> \kappa$}
\State {Update $P_m$ and $\mathcal{M}^{+}$ via \eqref{update_POWER};}
\State {Update $\mathbf{v}_{m}$ via Algorithm \ref{digt_algo_bf};}
\State {Update $\widetilde{R}_S^{\left(t+1\right)}$; $t = t+1$;}
\EndWhile
\State \textbf{Output}: {$\mathcal{M}^{+}$, $P_m$, $\mathbf{v}_{m}$.} 
\end{algorithmic}
\end{algorithm}

\subsection{Approximation via Cascaded TTD and PS}
After obtaining the semi-digital solution, we proceed to approximate it via exploiting the ATP structure. Specifically, by tuning the phased of the PSs and the delays of the TTDs, we aim to minimize the sum of the mean square errors between the semi-digital beamformers and analog ones across the subcarriers. The optimization problem is formulated as
\begin{subequations}\label{approx_ori}
\begin{align}
\mathop {\min }\limits_{\mathbf{w}, \tau_i} \;\;\; & \eta\buildrel \Delta \over=  \sum_{m \in \mathcal{M}^{+}} \left\|\mathbf{v}_{m}-\text{diag}\left(\mathbf{w}\right)\mathbf{t}_m\right\|^2  \label{approx_ori-obj}\\
{\rm{s.t.}}\;\;\;
&\eqref{ori_prob-ps},\eqref{ori_prob-ttd}.
\end{align}
\end{subequations}
The objective function of \eqref{approx_ori} can be converted into
\begin{equation}\label{approx_obj_tran}
\eta=2\text{card}\left(\mathcal{M}^{+}\right)N - 2\sum_{m \in \mathcal{M}^{+}} \text{Re}\left[\mathbf{v}_{m}^{H}\text{diag}\left(\mathbf{w}\right)\mathbf{t}_m\right].
\end{equation}
As a result, \eqref{approx_ori} is equivalent to
\begin{subequations}\label{approx_ori_eq}
\begin{align}
\mathop {\max }\limits_{\mathbf{w}, \tau_i} \;\;\; & \sum_{m \in \mathcal{M}^{+}} \text{Re}\left[\mathbf{v}_{m}^{H}\text{diag}\left(\mathbf{w}\right)\mathbf{t}_m\right]  \label{approx_ori_eq-obj}\\
{\rm{s.t.}}\;\;\;
&\eqref{ori_prob-ps},\eqref{ori_prob-ttd}.
\end{align}
\end{subequations}
Note that it is intractable to configure PSs and TTDs simultaneously due to the coupling. Again, we leverage the AO framework which optimizes $\mathbf{w}$ and $\tau_i$ alternatively.

\emph{1) Subproblem of Optimizing PS:} 
For given TTDs, the subproblem of optimizing PSs is recast into
\begin{subequations}\label{approx_ori_eq_ps}
\begin{align}
\mathop {\max }\limits_{\mathbf{w}} \;\;\; & \text{Re}\left[\left(\sum_{m \in \mathcal{M}^{+}}\mathbf{v}_{m}^{H}\text{diag}\left(\mathbf{t}_m\right)\right)\mathbf{w}\right]  \label{approx_ori_eq_ps-obj}\\
{\rm{s.t.}}\;\;\;
&\eqref{ori_prob-ps}.
\end{align}
\end{subequations}
It is straightforward to see that the optimal PS-based beamformer is
\begin{equation}\label{opt_w}
\mathbf{w}^{*} = e^{-\angle{\left(\sum_{m \in \mathcal{M}^{+}}\mathbf{v}_{m}^{H}\text{diag}\left(\mathbf{t}_m\right)\right)}}.  
\end{equation}

\emph{2) Subproblem of Optimizing TTD:} For given PSs, the subproblem of optimizing TTDs can be reformulated as
\begin{subequations}\label{approx_ori_eq_ttd}
\begin{align}
\mathop {\max }\limits_{\tau_i} \;\;\; & \sum_{m \in \mathcal{M}^{+}} \text{Re}\left[\sum_{i=1}^{N_T}\sum_{j=1}^{N_G}\psi_{m,\left(i-1\right)N_G+j} e^{-j2\pi f_m \tau_i} \right]  \label{approx_ori_eq_ttd-obj}\\
{\rm{s.t.}}\;\;\;
&\eqref{ori_prob-ttd},
\end{align}
\end{subequations}
where $\psi_{m,n}$ is the element at the $m$-th row and $n$-th column of matrix $\mathbf{\Psi}\in \mathbb{C}^{M\times N}$, whose $m$-th row is $\mathbf{v}_{m}^{H}\text{diag}\left(\mathbf{w}\right)$. Furthermore, we define $\gamma_{m,i,j}(\tau_i)\buildrel \Delta \over=\cos(2\pi f_m \tau_i-\zeta_{m,i,j})$, where $\zeta_{m,i,j}=\angle \psi_{m,(i-1)N_G+j}-\pi$. The part solely relying on $\tau_i$ can be separated from \eqref{approx_ori_eq_ttd} and rewritten as
\begin{subequations}\label{approx_ori_eq_ttd_sub}
\begin{align}
\mathop {\min }\limits_{\tau_i} \;\;\; &  \sum_{m \in \mathcal{M}^{+}}\sum_{j=1}^{N_G}\big|\psi_{m,\left(i-1\right)N_G+j}\big|\gamma_{m,i,j}\left(\tau_i\right)  \label{approx_ori_eq_ttd_sub-obj}\\
{\rm{s.t.}}\;\;\;
&0 \le \tau_i \le \chi . \label{approx_ori_eq_ttd_sub-budget}
\end{align}
\end{subequations}

Note that \eqref{approx_ori_eq_ttd_sub} is the sum of scaled cosine functions, which is non-convex. To circumvent this challenge, we turn to the BSUM optimization framework \cite{razaviyayn2013unified,zack2023twc} which successively finds a global upper bound on the objective function, ensuring it has the same value and derivative at a given point, and subsequently minimizes the upper bound. 
Using BSUM, we optimize $\tau_i$ as follows. In the $v$-th iteration, we define a quadratic function
\begin{equation}\label{quad func}
g^{\left(v\right)}_{m,i,j}\left(\tau_i\right)\buildrel \Delta \over =a^{\left(v\right)}_{m,i,j}\left(\tau_i-b^{\left(v\right)}_{m,i,j}\right)^2+c^{\left(v\right)}_{m,i,j},
\end{equation}
where $a^{\left(v\right)}_{m,i,j}$, $b^{\left(v\right)}_{m,i,j}$, and $c^{\left(v\right)}_{m,i,j}$ are chosen such that 
\begin{subequations}\label{quad coe condition}
\begin{align}
&g^{\left(v\right)}_{m,i,j}\left(\tau^{\left(v-1\right)}_i\right)=\gamma_{m,i,j}\left(\tau^{\left(v-1\right)}_i\right),\label{qcc-a}\\
&g^{\left(v\right)'}_{m,i,j}\left(\tau^{\left(v-1\right)}_i\right)=\gamma_{m,i,j}^{'}\left(\tau^{\left(v-1\right)}_i\right),\label{qcc-b}\\
&g^{\left(v\right)}_{m,i,j}\left(\tau_i\right) \ge \gamma_{m,i,j}\left(\tau_i\right),0 \le \tau_i \le \chi.\label{qcc-c}
\end{align}
\end{subequations}

\begin{figure}[!t]
\centering
\includegraphics[width=0.49\textwidth]{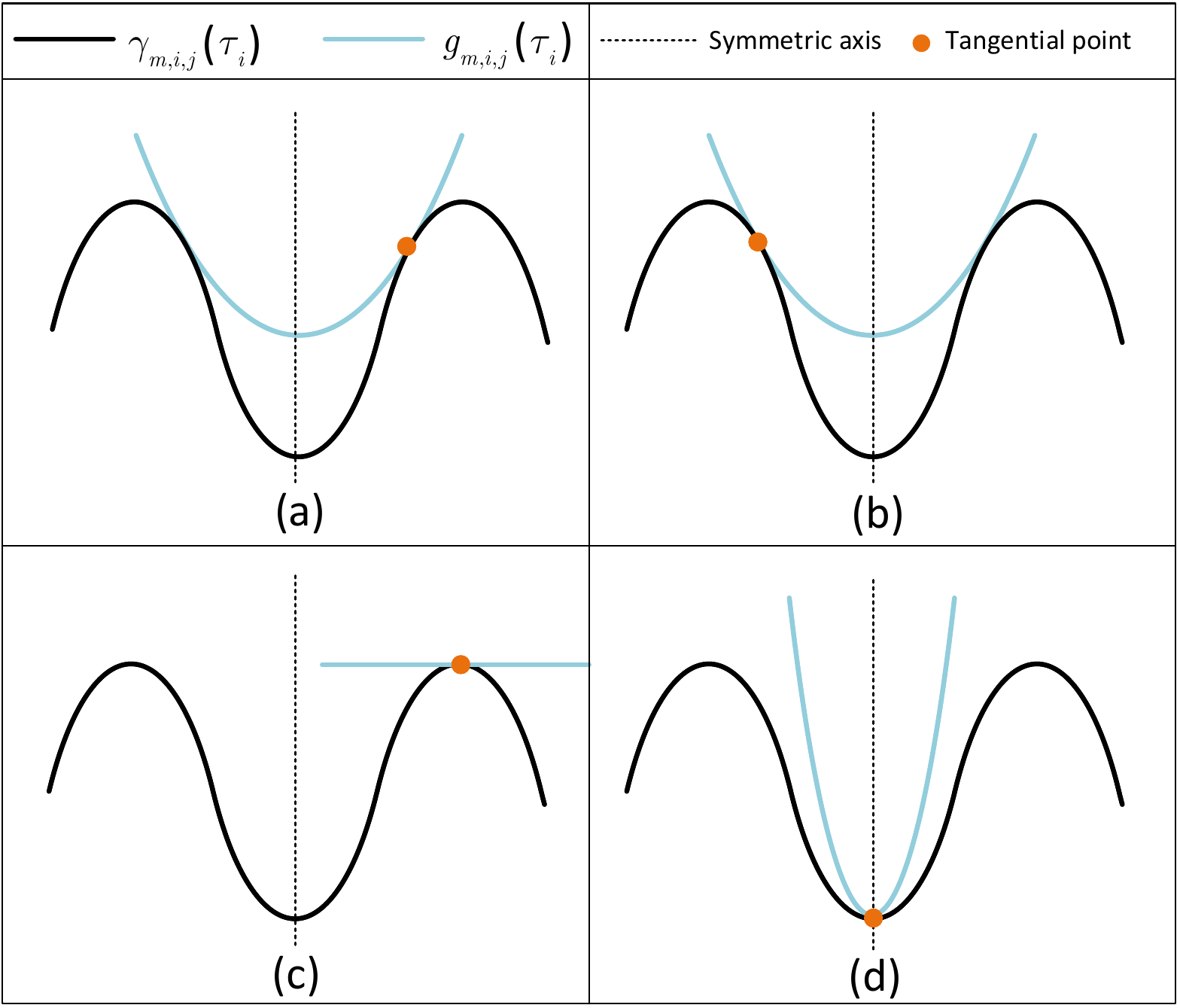}
\caption{Different cases of $g^{(v)}_{m,i,j}(\tau_i)$. (a) Case 1: $\gamma_{m,i,j}^{'}(\tau^{(v-1)}_i)>0$. (b) Case 2: $\gamma_{m,i,j}^{'}(\tau^{(v-1)}_i)<0$. (c) Case 3: $\gamma_{m,i,j}^{'}(\tau^{(v-1)}_i)=0$ and $\gamma_{m,i,j}(\tau^{(v-1)}_i)=1$. (d) Case 4: $\gamma_{m,i,j}^{'}(\tau^{(v-1)}_i)=0$ and $\gamma_{m,i,j}(\tau^{(v-1)}_i)=-1$.}
\label{bsum mod}
\end{figure} 

In order to meet the requirements in \eqref{qcc-a} and \eqref{qcc-b}, it is essential for both $g^{(v)}_{m,i,j}(\tau_i)$ and $\gamma_{m,i,j}(\tau_i)$ to be simultaneously tangential at $\tau_i = \tau^{(v-1)}_i$. Depending on the values of $\gamma_{m,i,j}^{'}(\tau^{(v-1)}_i)$ and $\gamma_{m,i,j}(\tau^{(v-1)}_i)$, we establish $g^{(v)}_{m,i,j}(\tau_i)$ in four cases to ensure the fulfillment of \eqref{qcc-c}:

{\bf{Case 1}}: When $\gamma_{m,i,j}^{'}(\tau^{(v-1)}_i)>0$, as illustrated in Fig. \ref{bsum mod}(a), we construct $g^{(v)}_{m,i,j}(\tau_i)$ such that it exhibits symmetry around the first left valley point of $\gamma_{m,i,j}(\tau_i)$ adjacent to the tangential point.

{\bf{Case 2}}: In the event that $\gamma_{m,i,j}^{'}(\tau^{(v-1)}_i)<0$, as depicted in Fig. \ref{bsum mod}(b), we design $g^{(v)}_{m,i,j}(\tau_i)$ to be symmetrical around the first right valley point of $\gamma_{m,i,j}(\tau_i)$ next to the tangential point.

{\bf{Case 3}}: When $\gamma_{m,i,j}^{'}(\tau^{(v-1)}_i)=0$ and $\gamma_{m,i,j}(\tau^{(v-1)}_i)=1$, as shown in Fig. \ref{bsum mod}(c), we set $g^{(v)}_{m,i,j}(\tau_i)=1$.

{\bf{Case 4}}: In the situation where $\gamma_{m,i,j}^{'}(\tau^{(v-1)}_i)=0$ and $\gamma_{m,i,j}(\tau^{(v-1)}_i)=-1$, as presented in Fig. \ref{bsum mod}(d), we construct $g^{(v)}_{m,i,j}(\tau_i)$ to be symmetrical about $\tau_i = \tau^{(v-1)}_i$. Additionally, we ensure that the second-order derivatives of $g^{(v)}_{m,i,j}(\tau_i)$ and $\gamma_{m,i,j}(\tau_i)$ match at $\tau_i = \tau^{(v-1)}_i$, which can be easily verified as a sufficient condition for \eqref{qcc-c} to hold.

In light of the aforementioned cases, the values of $a^{\left(v\right)}_{m,i,j}$, $b^{\left(v\right)}_{m,i,j}$, and $c^{\left(v\right)}_{m,i,j}$ can be derived as follows.

\begin{itemize}
\item When $\gamma_{m,i,j}^{'}(\tau^{(v-1)}_i)\neq 0$,
\begin{subequations}\label{quad coe case 1}
\begin{align}
a^{\left(v\right)}_{m,i,j}=&\frac{-\pi f_m \sin \left(2\pi f_m \tau_i^{\left(v-1\right)}-\zeta_{m,i,j}\right)}{\tau_i^{\left(v-1\right)}-b^{\left(v\right)}_{m,i,j}},\\
b^{\left(v\right)}_{m,i,j}=&
\left\{ {\begin{array}{*{20}{l}}
&\frac{\lfloor 2f_m \tau_i^{\left(v-1\right)}-\frac{\zeta_{m,i,j}}{\pi} \rfloor + \frac{\zeta_{m,i,j}}{\pi}}{2f_m},\\
&\gamma_{m,i,j}^{'}\left(\tau^{\left(v-1\right)}_i\right)>0, \\
&\frac{\lceil 2f_m \tau_i^{\left(v-1\right)}-\frac{\zeta_{m,i,j}}{\pi} \rceil + \frac{\zeta_{m,i,j}}{\pi}}{2f_m},\\
&\text{otherwise},\\
\end{array}} 
\right.\\
c^{\left(v\right)}_{m,i,j}=&\cos \left(2\pi f_m \tau_i^{\left(v-1\right)}-\zeta_{m,i,j}\right) \nonumber\\
&\;\;\;\;\;\;\;\;\;\;\;\;-b^{\left(v\right)}_{m,i,j}\left(\tau_i^{\left(v-1\right)}-a^{\left(v\right)}_{m,i,j}\right)^2.	
\end{align}	
\end{subequations}

\item Otherwise,
\begin{subequations}\label{quad coe case 2}
\begin{align}
&a^{\left(v\right)}_{m,i,j}=\left\{ {\begin{array}{*{20}{l}}
0,&\gamma_{m,i,j}\left(\tau_i^{\left(v-1\right)}\right)=1,\\
2\pi^2f_m^2,&\text{otherwise},\\
\end{array}} \right.	\\
&b^{\left(v\right)}_{m,i,j}=\tau_i^{\left(v-1\right)},\\
&c^{\left(v\right)}_{m,i,j}=\left\{ {\begin{array}{*{20}{l}}
1,&\gamma_{m,i,j}\left(\tau_i^{\left(v-1\right)}\right)=1,\\
-1,&\text{otherwise}.\\
\end{array}} \right.	
\end{align}
\end{subequations}
\end{itemize}

As a result, \eqref{approx_ori_eq_ttd_sub} is replaced by   
\begin{subequations}\label{approx_ori_eq_ttd_sub_bsum}
\begin{align}
\mathop {\min }\limits_{\tau_i} \;\;\; & \sum_{m \in \mathcal{M}^{+}}\sum_{j=1}^{N_G}\big|\psi_{m,\left(i-1\right)N_G+j}\big|g^{\left(v\right)}_{m,i,j}\left(\tau_i\right)  \label{approx_ori_eq_ttd_sub_bsum-obj}\\
{\rm{s.t.}}\;\;\;
&\eqref{approx_ori_eq_ttd_sub-budget}.
\end{align}
\end{subequations}
The solution to \eqref{approx_ori_eq_ttd_sub_bsum} is then given to $\tau_i^{(v)}$. The objective function of \eqref{approx_ori_eq_ttd_sub_bsum} is a summation of convex quadratic functions. Consequently, it bears the form of a continuous convex quadratic function with respect to $\tau_i \in [0, \chi]$. Therefore, we can readily express the unique optimal solution to \eqref{approx_ori_eq_ttd_sub_bsum} as
\begin{equation}\label{opt_tau}
\tau_i^{\left(v\right)}=
\left\{ {\begin{array}{*{20}{l}}
0,
&\widetilde{\tau}_i^{\left(v\right)}<0,\\
\widetilde{\tau}_i^{\left(v\right)},
&0 \le \widetilde{\tau}_i^{\left(v\right)} \le  \chi,\\
\chi,
&\text{otherwise},
\end{array}} \right.
\end{equation}
where 
\begin{equation}\label{opt_tau_close}
\widetilde{\tau}_i^{\left(v\right)}=\frac{\sum_{m \in \mathcal{M}^{+}}\sum_{j=1}^{N_G}\big|\psi_{m,\left(i-1\right)N_G+j}\big|a^{\left(v\right)}_{m,i,j}b^{\left(v\right)}_{m,i,j}}{\sum_{m \in \mathcal{M}^{+}}\sum_{j=1}^{N_G}\big|\psi_{m,\left(i-1\right)N_G+j}\big|a^{\left(v\right)}_{m,i,j}}.    
\end{equation}
As a result, \eqref{approx_ori_eq_ttd} can be solved by following the updating rules specified above in an iterative manner.

\renewcommand{\algorithmicrequire}{\textbf{Input:}}
\renewcommand{\algorithmicensure}{\textbf{Output:}}
\begin{algorithm}[t]
\caption{TTD Optimization Based on BSUM}
\label{algo_bsum}       %
\begin{algorithmic}[1]
\State \textbf{Input}: $\mathcal{M}^{+}$, $\mathbf{v}_{m}$, $\vartheta$;
\State \textbf{Initialize}: $v=0$, $\boldsymbol{\tau}^{\left(0\right)}$;
\While {$\left\|\boldsymbol{\tau}^{(v+1)}-\boldsymbol{\tau}^{(v)}\right\|>\vartheta$;}
\For {$i\in \mathcal{T}$, $m\in \mathcal{M}^{+}$, $j\in \mathcal{G}\buildrel \Delta \over=\{1,\ldots,N_G\}$}
\State Update $a^{(v+1)}_{m,i,j}$, $b^{(v+1)}_{m,i,j}$, $c^{(v+1)}_{m,i,j}$, via \eqref{quad coe case 1} and \eqref{quad coe case 2};
\State Update $\tau_i^{(v+1)},\forall i\in \mathcal{T}$, via \eqref{opt_tau}; 
\EndFor  { $v=v+1$;}
\EndWhile
\State  Output $\tau_i$.
\end{algorithmic}
\end{algorithm}

Algorithm \ref{algo_bsum} details the steps involved in optimizing TTDs using BSUM, where $\boldsymbol{\tau}^{(v)}\buildrel \Delta \over=[\tau_1^{\left(v\right)},\ldots,\tau_{N_T}^{\left(v\right)}]^T \in \mathbb{C}^{N_T \times 1}$ and $\vartheta$ is the iterative tolerance. The convergence of this algorithm can be readily confirmed by examining the conditions outlined in Theorem 2 of \cite{razaviyayn2013unified}. Each limit point produced by Algorithm \ref{algo_bsum} corresponds to a stationary point of the \eqref{approx_ori_eq_ttd}. The overall configuration of the ATP structure via AO is presented in Algorithm \ref{algo_atp}, where $\varsigma$ is the iterative tolerance of AO.
The computational complexity of Algorithm \ref{algo_atp} per iteration is dominated by Algorithm \ref{algo_bsum}, which is $\mathcal{O}(N_T)$.

\renewcommand{\algorithmicrequire}{\textbf{Input:}}
\renewcommand{\algorithmicensure}{\textbf{Output:}}
\begin{algorithm}[t]
\caption{Analog Beamformer Design via AO}
\label{algo_atp}       %
\begin{algorithmic}[1]
\State \textbf{Input}: $\mathcal{M}^{+}$, $\mathbf{v}_{m}$, $\varsigma$;
\State \textbf{Initialize}: $l=0$, $\boldsymbol{\tau}^{\left(0\right)}$, $\eta^{\left(0\right)}$;
\While {$|\eta^{\left(l+1\right)}-\eta^{\left(l\right)}|> \varsigma$}
\State {Update $\mathbf{w}$ via \eqref{opt_w};}
\State {Update $\tau_i$ via Algorithm \ref{algo_bsum};}
\State {Update $\eta^{\left(l+1\right)}$; $l = l+1$;}
\EndWhile
\State \textbf{Output}: {$\mathbf{w}$, $\tau_i$.} 
\end{algorithmic}
\end{algorithm}

The analog wideband beamfocusing can be realized by sequentially executing Algorithms \ref{digt_algo} and \ref{algo_atp}. However, due to the limited capability of analog devices, there could be a mismatch between the semi-digital and analog beamformers, therefore changing the indexes of carriers which power should be allocated to. To prevent potential power wastage, $\mathcal{M}^{+}$ is supposed to be updated with the analog beamformer synthesized as $\text{diag}\left(\mathbf{w}\right)\mathbf{t}_m$. Moreover, the power allocation should be refined as per \eqref{opt_pow}.

\section{Beamsplit-Aware Low-Complexity Approach}
While our proposed beamformer in the last section ensures secure wideband beamfocusing in the near-field region, it has two major challenges. On the one hand, the convex optimization imposes significant computational burden as the complexity is proportional to the antenna number. On the other hand, the overall performance is sensitive to the initial values of delay in Algorithm \ref{algo_atp}, emphasizing the importance of carefully choosing $\boldsymbol{\tau}^{\left(0\right)}$ for robust performance. In this section, we address these issues by presenting a beamsplit-aware low-complexity approach (BALA) which exploits the geometric property inherent in the near-field wideband propagation to achieve efficient analog beamfocusing. Moreover, the corresponding delays configured by this approach can be tailored to serve as initial values for Algorithm \ref{algo_atp}, ensuring robustness.

\begin{figure}[!t]
\centering
\includegraphics[width=0.5\textwidth]{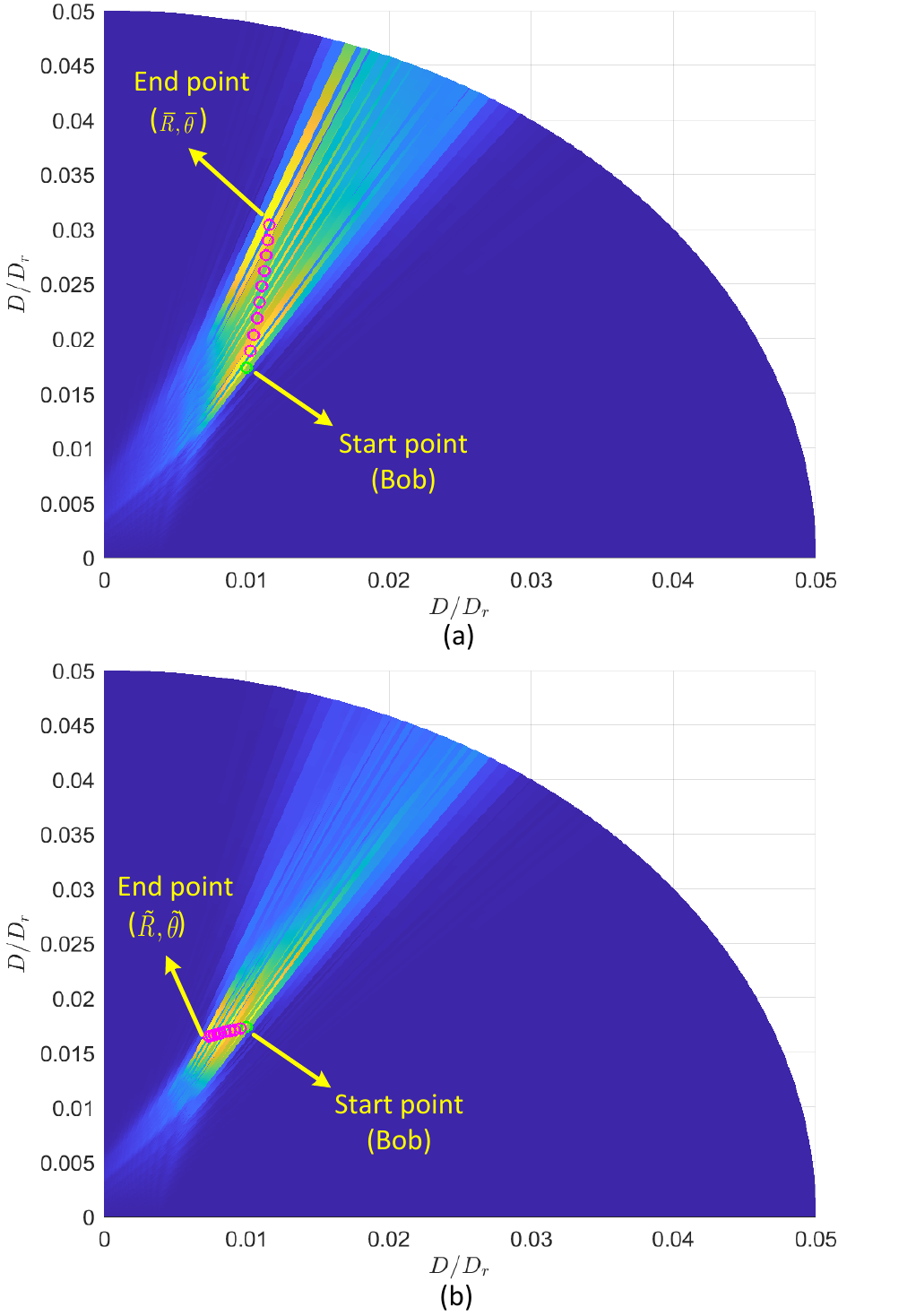}
\caption{Synthesized wideband beampattern with axes normalized by Rayleigh distance $D_r$. (a) Without TTD. (b) With TTD. }
\label{bp-BALA}
\end{figure}

Without the assist of TTDs, we consider the analog beamformer is matched to frequency $f_1$ and the location of Bob, i.e., $\mathbf{w}_B=[e^{\frac{-j2\pi f_1 D_{B,1}}{c}},\ldots,e^{\frac{-j2\pi f_1 D_{B,N}}{c}}]^T$, where $D_{B,n}$ is the distance from the $n$-th antenna of Alice to Bob. As a consequence, the component of the wideband signal at frequency $f_1$ can be precisely focused at Bob, while other components deviate to different positions due to mismatches, creating a beamsplit effect.
The following proposition, which was proved in \cite{feifei2022arxiv}, presents the trace of the focus point as a function of frequency. 
\begin{proposition}\label{trace_prop_1}
Given $\mathbf{w}_B$, the near-field focus position at frequency $f$ can be approximated as 
\begin{subequations}\label{trace_without_ttd}
\begin{align}
& \theta = \arccos \left(\frac{f_1}{f}\cos\theta_B\right),\label{angle_without_ttd} \\
& R = \left(\frac{f}{f_1 \sin^2\theta_B}-\frac{f_1}{f \tan^2\theta_B}\right)R_B. \label{distance_without_ttd}
\end{align}
\end{subequations}
\end{proposition}
By examining Proposition \ref{trace_prop_1}, it becomes evident that as the carrier frequency increases, the deviation from Bob also increases, resulting in a significant loss in beamfocusing. The furthest deviated point, denoted by $(\overline{R},\overline{\theta})$, corresponds to $f = f_M$. To facilitate understanding, we illustrate the near-field TTD-free wideband beampattern in Fig. \ref{bp-BALA}(a), where the trace of focus points is determined by Proposition \ref{trace_prop_1}.
Intuitively, to improve the SR, it is preferable that the wideband signal components at each frequency be as focused as possible at the legitimate receiver Bob.
To mitigate detrimental misfocusing, we adopt the ATP structure, as introduced in Section II. Assuming the use of $N$ TTDs, i.e., the fully-connected structure, we provide the configuration of the ATP structure and the corresponding trace of the focus points in Proposition \ref{trace_with_ttd}. The derivation process mirrors that outlined in Section IV-B of \cite{feifei2022arxiv}. To maintain conciseness, we omit the details here.
\begin{proposition}\label{trace_with_ttd}
To focus the signal component at $f_1$ towards $(R_B,\theta_B)$ and the one at $f_M$ towards a specified point $(\widetilde{R},\widetilde{\theta})$, the settings for the $n$-th TTD and the $n$-th PS are
\begin{equation}\label{ttd_config}
\tau_n  = \frac{f_M \widetilde{D}_n - f_1 D_{B,n}}{cB}
\end{equation}
and
\begin{equation}\label{ps_config}
\phi_n  = \frac{2\pi f_1 f_M \left( \widetilde{D}_{n} - D_{B,n} \right)}{cB}, \forall n \in \mathcal{N},  
\end{equation}
respectively, where $\widetilde{D}_{n}$ is the distance from the $n$-th antenna of Alice to point $(\widetilde{R},\widetilde{\theta})$. The corresponding near-field focus position at frequency $f$ can be approximated as
\begin{subequations}
\begin{align}
& \theta = \arccos \left(\frac{\left(f-f_1\right)f_M \cos \widetilde{\theta} - \left(f-f_M\right)f_1 \cos \theta_B}{Bf}\right),\label{angle_with_ttd} \\
& R = 1 \bigg/\left(\frac{\left(f-f_1\right)f_M \sin^2\widetilde{\theta}}{B f \widetilde{R} \sin^2\theta}-\frac{\left(f-f_M\right)f_1 \sin^2\theta_B}{B f R_B \sin^2\theta }\right). \label{distance_with_ttd}
\end{align}
\end{subequations}
\end{proposition}
According to Proposition \ref{trace_with_ttd}, we can manipulate the end point of the trace via setting the ATP structure as per \eqref{ttd_config} and \eqref{ps_config}, thereby reducing the beamsplit effect.
The corresponding wideband beampattern is illustrated in Fig. \ref{bp-BALA}(b), where the trace of focus points is determined by Proposition \ref{trace_with_ttd}. The basic idea of BALA is to perform line search between the TTD-free end point $(\overline{R},\overline{\theta})$ and $(R_B,\theta_B)$, thus identifying the position with highest SR.
It should be emphasized that different delays are required as $(\widetilde{R},\widetilde{\theta})$ goes from $(\overline{R},\overline{\theta})$ to $(R_B,\theta_B)$, which could potentially violate the constraint for delay budget, i.e., $0 \le \tau_n \le \chi$. Meanwhile, to enhance the SR, we should not only mitigate the misfocus but also to minimize the energy leakage towards Eve. Therefore, once the configuration of the ATP structure is determined, the power across different carriers is supposed to be reallocated as per \eqref{update_POWER}. The procedure of BALA is outlined in Algorithm \ref{algo_bala}, with $(\overline{X},\overline{Y})$ and $(X_B,Y_B)$ denoting the Cartesian coordinates corresponding to the polar coordinates $(\widetilde{R},\widetilde{\theta})$ and $(R_B,\theta_B)$, respectively. Additionally, $L$ represents the number of search segments while $R_S^{temp}$ denotes the temporary register for SR.

\renewcommand{\algorithmicrequire}{\textbf{Input:}}
\renewcommand{\algorithmicensure}{\textbf{Output:}}
\begin{algorithm}[t]
\caption{Beamsplit-Aware Low-Complexity Approach}
\label{algo_bala}       %
\begin{algorithmic}[1]
\State \textbf{Input}: $(\overline{X},\overline{Y})$, $(X_B,Y_B)$, $L$;
\State \textbf{Initialize}: $R_S = 0$, $R_S^{temp} = 0$;
\For {$l = 1 : L$}
\State {Update $(\widetilde{X},\widetilde{Y})$ via $(\overline{X} + \frac{l(X_B-\overline{X})}{L},\overline{Y} + \frac{l(Y_B-\overline{Y})}{L})$;}
\State {Convert $(\widetilde{X},\widetilde{Y})$ into $(\widetilde{R},\widetilde{\theta})$;}
\State {Update $\tau_n$ and $\phi_n$ via \eqref{ttd_config} and \eqref{ps_config}, respectively;}
\State {Refine $\tau_n$ via $\max(0,\min(\tau_n,\chi))$;}
\State {Allocate power as per \eqref{update_POWER} and update $R_S^{temp}$;}
\State {Update $R_S$ via $R_S^{temp}$ and store $\tau_n$ and $\phi_n$ only when $R_S^{temp}> R_S$;}
\EndFor
\State \textbf{Output}: {$\tau_n$, $\phi_n$.} 
\end{algorithmic}
\end{algorithm}

When the number of TTDs is smaller than that of the antennas, i.e., $N_T \le N$, we should further refine, after Step 7 in Algorithm \ref{algo_bala}, the value of the $i$-th delay, $\forall i \in \mathcal{T}$, as the mean value of the delays indexed by $n \in \{(i-1)N_g+1,\ldots,i N_g\}$. The proposed BALA not only achieves low-complexity near-field wideband secure beamfocusing but also provides insights into configuring initial delays for Algorithm \ref{algo_atp}. 

\section{Numerical Results}
In this section, we present numerical results to validate the effectiveness of the proposed approaches for near-field wideband secure beamfocusing. Without otherwise specified, the basic simulation parameters are set as follows. The considered wideband system operates at a central carrier frequency $f_c = 24$ GHz, with a signal bandwidth $B= 8$ GHz and a total number of subcarriers $M=10$. The transmitter Alice is equipped with a 64-antenna ULA with half-central-wavelength interval. The Rayleigh distance can be calculated as $D_r =2A^2/\lambda \approx 50\text{ m}$, where the array aperture $A=(N-1)\lambda_c/2$ with $\lambda_c$ being the wavelength corresponding to central carrier frequency $f_c$. To construct the ATP structure, the system employs a single RF chain followed by 32 TTDs and 64 dedicated PSs connected to each of the antennas. We assume that Bob and Eve are located at $(0.02D_r,60^{\circ})$ and $(0.015D_r,65^{\circ})$, respectively. The transmit power budget and noise power spectrum density at each receiver are set as $20$ dBm and $-100$ dBm/Hz, respectively. The delay budget is $5$ ns. We also set iterative tolerances with $\delta=10^{-3}$, $\epsilon=10^{-3}$, $\kappa=10^{-3}$, $\vartheta=10^{-4}$, and $\varsigma=10^{-4}$. Additionally, we configure $L$ to be 100.


\begin{figure*}
\centering
\includegraphics[width=7 in]{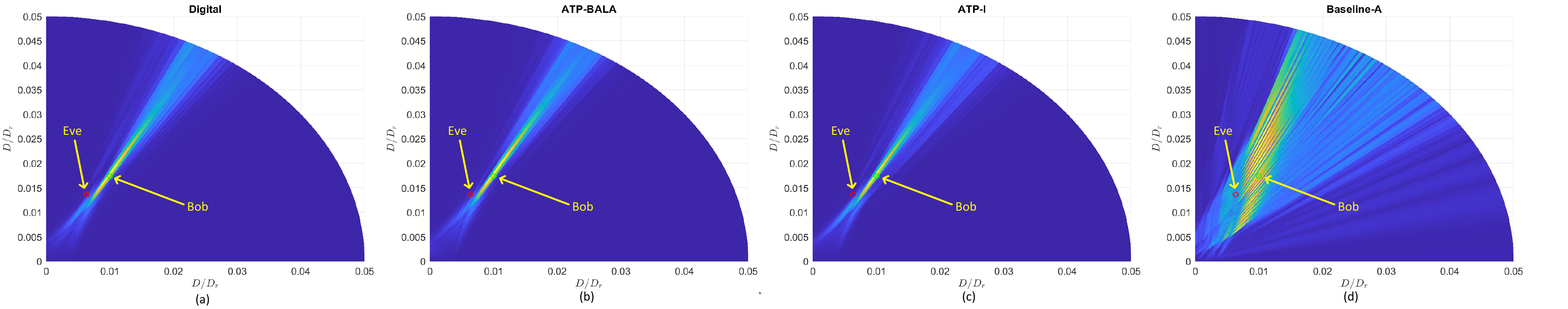}
\vspace{-6mm}
\caption{Synthesized wideband beampattern with axes normalized by Rayleigh distance $D_r$: (a) Fully-digital; (b) ATP-BALA; (c) ATP-I; (d) Baseline-A.}
\label{bp_collect}	
\end{figure*}

In our two-stage optimization framework, we compare the semi-digital problem \eqref{digit_prob} with its fully-digital counterpart, which is solved in a similar manner except for replacing constraint \eqref{digit_prob-bf} with $\big\|\mathbf{v}_m\big\| = \sqrt{N},\forall m \in \mathcal{M}$. We label optimization-based beamfocusing approaches with first-stage semi-digital and fully-digital methods as Scheme-I and Scheme-II, respectively.
We denote Algorithm \ref{algo_atp} with first-stage optimization as Scheme-I and Scheme-II ATP-I and ATP-II, respectively. Additionally, we refer Algorithm \ref{algo_bala} as ATP-BALA.

To explore the role of TTD, we introduce two TTD-free analog modifications as benchmarks: 

Baseline-A: It employs Algorithm \ref{algo_atp} with all delays set to 0, eliminating the need for AO of PSs and TTDs. The desired beamformer is then obtained via \eqref{opt_w}, followed by power allocation according to \eqref{update_POWER}, originally from \cite{eduard2008globecom}. 

Baseline-B: It simply sets the analog beamformer as the normalized channel gain at frequency $f_1$ and allocates power across subcarriers using \eqref{update_POWER}, originally from \cite{eduard2008globecom}.

Additionally, we compare a representative TTD-free analog secure beamforming benchmark designed for narrowband systems \cite{analogboyd2020}, referred as Baseline-C. We have adapted this approach for the wideband context.

\begin{figure*}
\centering
\includegraphics[width=6.6 in]{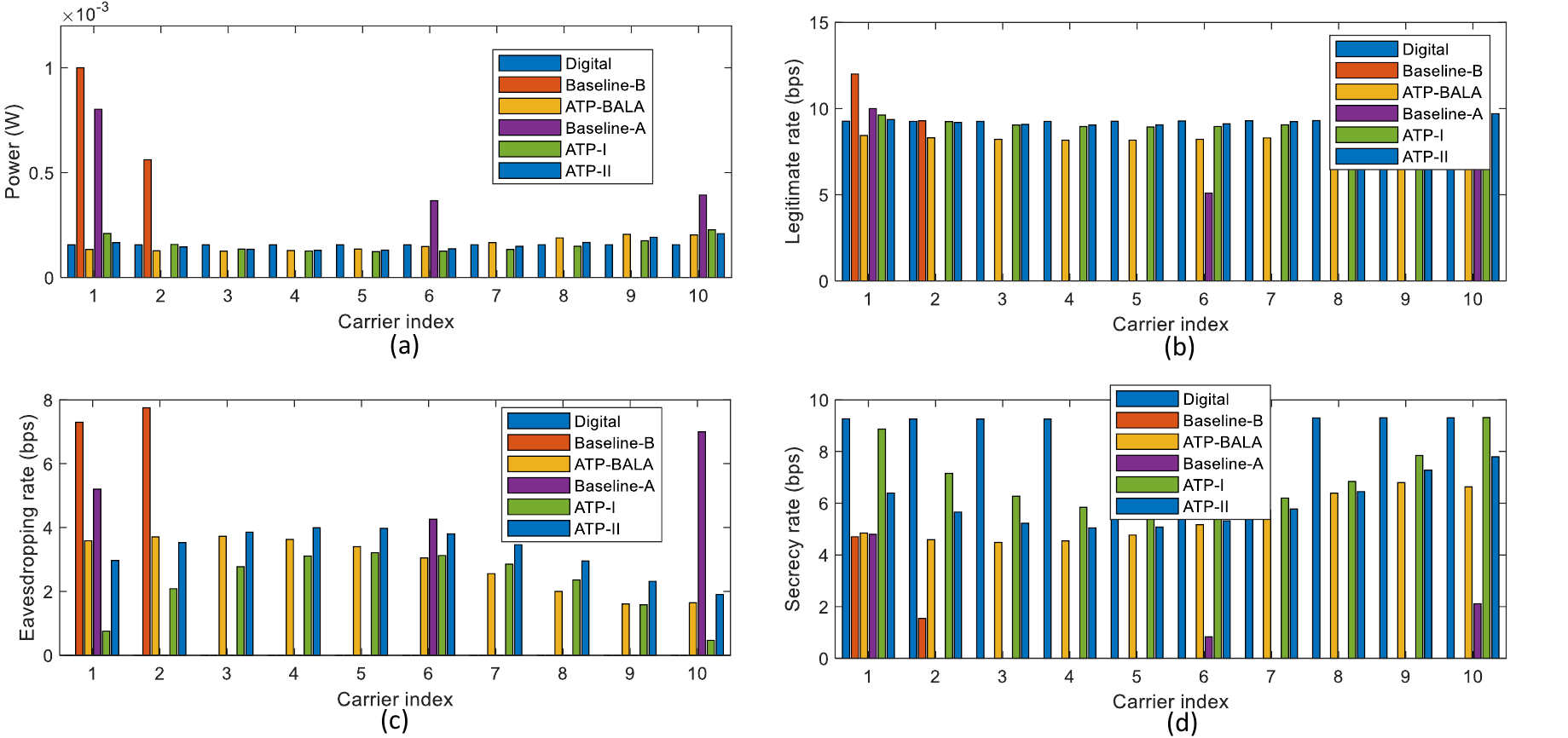}
\vspace{-4mm}
\caption{Per-subcarrier power/rate evaluation: (a) Power; (b) Legitimate rate at Bob; (c) Eavesdropping rate at Eve; (d) Secrecy rate.}
\label{resource_bar}	
\end{figure*}

Figs. \ref{bp_collect}(a)-(d) depict the synthesized wideband beampatterns, which result from superposing the beampatterns corresponding to each subcarrier using approaches: Fully-digital, ATP-I, ATP-BALA, and Baseline-A, respectively. 
The ones corresponding to ATP-II and Baseline-B are similar to that of ATP-I and Baseline-A, and are therefore omitted for conciseness. 
As observed, with the assistance of TTDs, all ATP approaches achieve significantly more focused beampatterns at Bob which are comparable to that achieved by the fully-digital solution. This effectively reduces energy leakage towards Eve and results in more favorable channel conditions at each subcarrier for secure transmission. This point will be further discussed in the subsequent results about subcarrier power allocation and corresponding rates.

Figs. \ref{resource_bar}(a)-(d) present a comparison of six different approaches, showcasing per-subcarrier power allocation, legitimate rates at Bob, eavesdropping rates at Eve, and secrecy rates, respectively. 
As shown in Fig. \ref{resource_bar}(a), the two TTD-free approaches selectively allocate power to only a few subcarriers with better channel quality towards Bob. In contrast, the ATP-assisted approaches utilize all subcarriers and allocate power more or less equally because, in principle, all corresponding channels exhibit good performance, owing to the rather focused beampattern illustrated in Figs. \ref{resource_bar}. This phenomenon underscores the unique benefits enabled by TTDs. Consequently, the ATP-assisted approaches effectively utilize the near-field wideband channel, delivering favorable performance across all subcarriers. This, in turn, contributes to significantly enhanced secrecy rates, as depicted in Fig. \ref{resource_bar}(d). Furthermore, when compared to the heuristic geometry-inspired approach ATP-BALA, both ATP-I and ATP-II demonstrate superior performance due to their ability to refine ATP-BALA's results through optimization techniques. In fact, Fig. \ref{resource_bar}(c) illustrates that this refinement further reduces energy leakage towards Eve, as evidenced by decreased eavesdropping rates in each subcarrier under ATP-I and ATP-II.

\begin{figure}
\centering
\includegraphics[width=3 in]{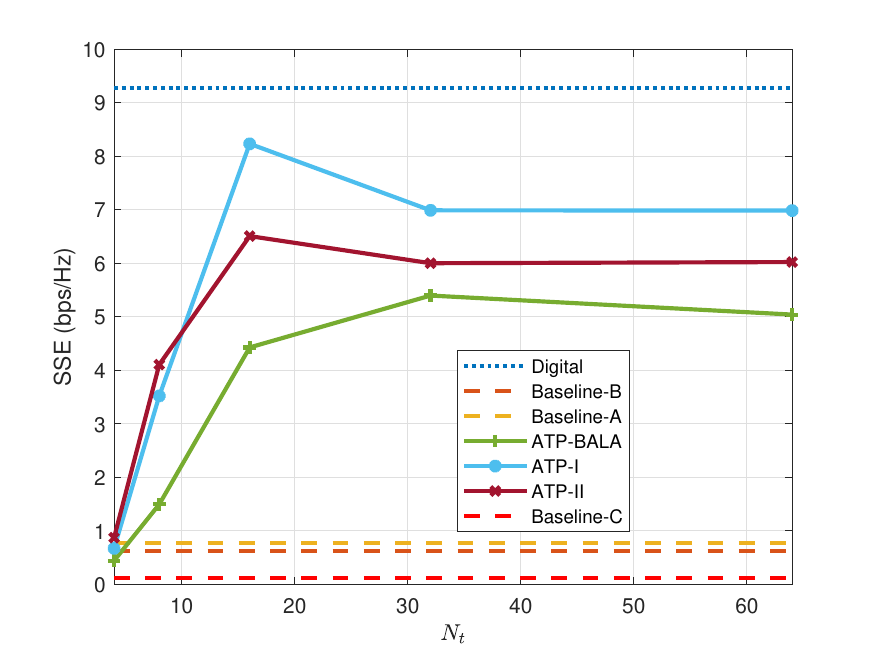}
\caption{SSE versus the number of TTDs.}\label{SE-Nt}	
\end{figure}

\begin{figure}
\centering
\includegraphics[width=3 in]{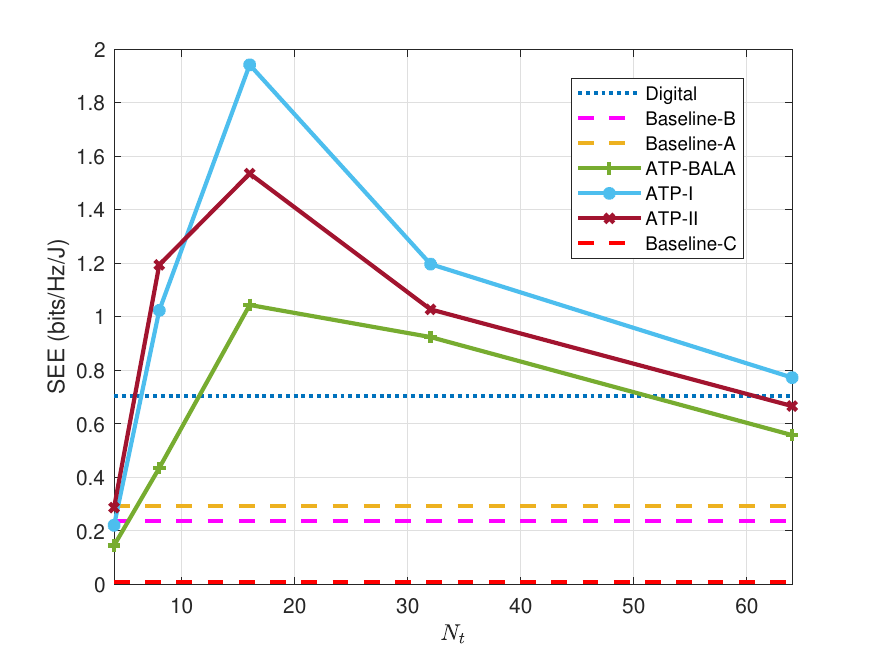}
\caption{SEE versus number of TTDs.}\label{EE-Nt}	
\end{figure}

Fig. \ref{SE-Nt} depicts the comparison of secrecy spectral efficiency (SSE), which is defined as the ratio of secrecy rate to bandwidth, with the number of TTD. Generally, the SSE of the three ATP-based approaches exhibits an upward trend with an increasing number of TTDs, outperforming the TTD-free approach. However, slight fluctuations occur due to the heuristic nature of the initial values.
When comparing ATP-BALA to other ATP-based approaches, it becomes apparent that the latter achieves higher SSE, affirming the effectiveness of the proposed AO-based configuration for the ATP structure. 
Additionally, Scheme-I surpasses Scheme-II. This superiority stems from the semi-digital first-stage optimization in Scheme-I, which enforces unit-modulus solutions. Consequently, it leads to a superior second-stage analog approximation with smaller mismatches. This clearly illustrates the advantage of selecting semi-digital optimization over fully-digital approaches in the first-stage optimization of the proposed two-stage framework.


Fig. \ref{EE-Nt} illustrates the corresponding secrecy energy efficiency (SEE), which is defined as the ratio of SSE to total power consumption, for different approaches, where the total power consumption accounts for the transmit power, baseband processing power $P_{\text{BB}}=25$ dBm, power consumption of RF chains $N_{\text{RF}}P_{\text{RF}}$, TTDs $N_{T}P_{\text{TTD}}$, and PSs $N P_{\text{PS}}$. Here, $N_{\text{RF}}$ represents the number of RF chains, and power consumption per RF chain, TTD, and PS are denoted as $P_{\text{RF}}=23$ dBm, $P_{\text{TTD}}=20$ dBm, and $P_{\text{PS}}=15$ dBm, respectively. As observed, the proposed ATP-based analog beamfocusing approaches exhibit a significant advantage over their digital counterpart in enhancing SEE across a broad range of TTD numbers, while maintaining a tolerable SSE degradation, as demonstrated in Fig. \ref{SE-Nt}. Moreover, Fig. \ref{SE-Nt} reveals that achieving peak SSE only necessitates a moderate number of TTDs, resulting in exceptionally high SEE with a small number of TTDs. However, as the number of TTDs further increases, SEE starts to decrease due to diminishing marginal returns.
These observations can be leveraged to reduce hardware costs in practice and enhance energy efficiency while maintaining a high secrecy rate.

\begin{figure}
\centering
\includegraphics[width=3 in]{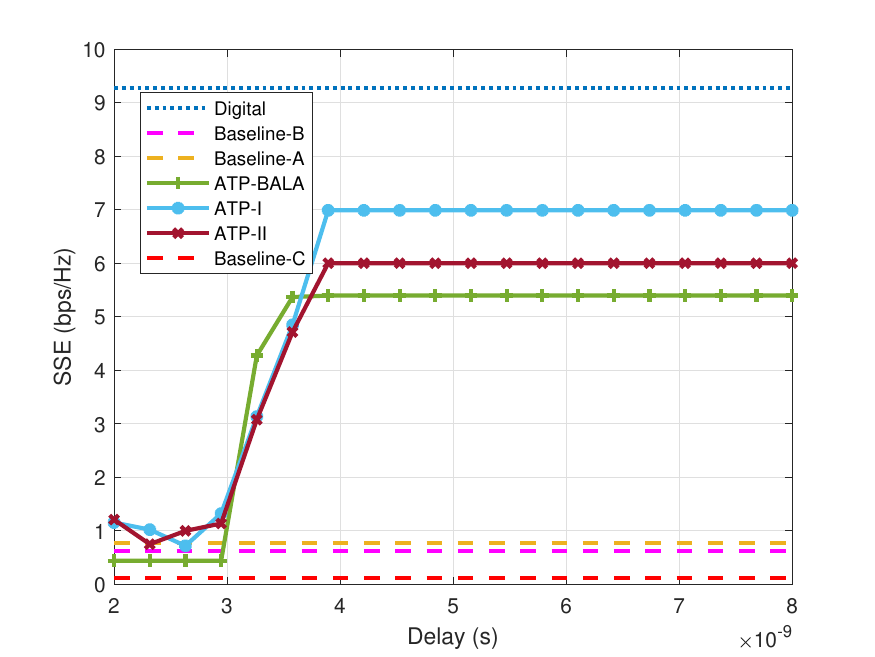}
\caption{SSE versus the delay budget.}\label{SE-Delay}	
\end{figure}


Fig. \ref{SE-Delay} illustrates the SSE versus the delay budget for various approaches. Notably, in cases of extremely limited delay budgets, the ATP-BALA approach performs worse than the TTD-free alternatives. This disparity arises because the constrained delay budget significantly alters the delays calculated by \eqref{ttd_config} during Step 7 of Algorithm \ref{algo_bala}, leading to a beampattern vastly different from the one promised by Proposition \ref{trace_with_ttd}. 
For the same underlying reason, the other ATP-based approaches also do not exhibit superior performance due to their dependence on a poor initial value. As the delay budget gradually increases, the SSE of ATP-assisted approaches improves, with optimization-refined methods displaying a clear advantage. Remarkably, ATP-I eventually reaches a level with only moderate degradation compared to the fully-digital solution. The SSE plateaus after a certain point. Hence, it is essential to make judicious decisions regarding the delay budget since excessive values provide no benefit and only contribute to delays.

\begin{figure}
\centering
\includegraphics[width=3 in]{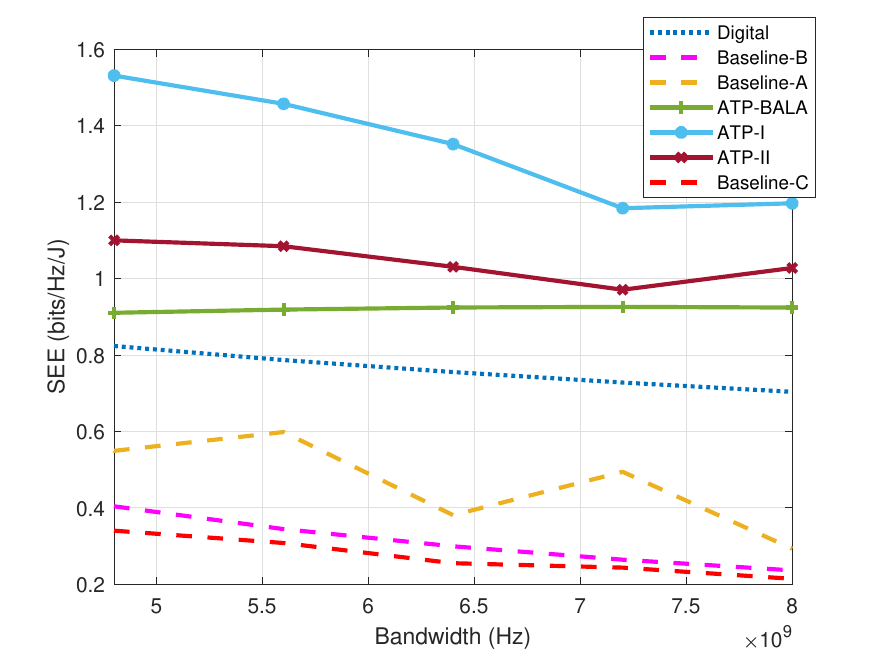}
\caption{SEE versus bandwidth.}\label{EE-bw}	
\end{figure}

\begin{figure}
\centering
\includegraphics[width=3 in]{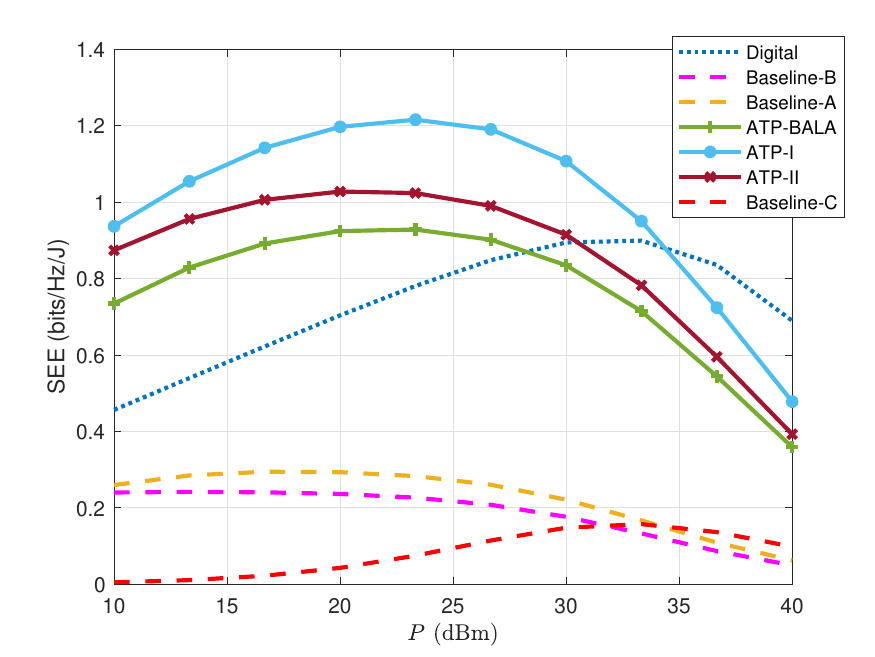}
\caption{SEE versus transmit power.}\label{EE-power}	
\end{figure}


Fig. \ref{EE-bw} examines SEE versus bandwidth across various approaches. The advantage of the proposed ATP-assisted methods in enhancing SEE is clearly evident, with ATP-based approaches displaying distinct superiority over fully-digital and TTD-free counterparts. Regardless of TTD availability, SEE generally decreases as bandwidth increases. This phenomenon occurs due to the larger bandwidth leading to a more pronounced near-field beamsplit. Consequently, mitigating this effect and achieving a focused beam pattern toward Bob, thereby enhancing security, necessitates a greater number of TTDs.

Fig. \ref{EE-power} compares the SEE of the proposed schemes concerning transmit power. At lower transmit power levels, increasing power significantly enhances secrecy rate and, consequently, SEE. However, as transmit power rises, the marginal improvement in secrecy rate diminishes, leading to a decline in SEE. The proposed ATP-assisted methods demonstrate superior SEE compared to their digital counterpart. Across a wide range of transmit power values, the proposed ATP-assisted beamfocusing schemes outperforms digital solutions, thanks to the substantial power savings achieved by utilizing more cost-effective analog devices instead of power-intensive RF chains.

\section{Conclusion}
This paper addressed the critical need for enhancing PLS in near-field wideband communications within the evolving 6G context. We introduced analog beamfocusing techniques utilizing TTDs and PSs to effectively mitigate beamsplit and exploit near-field advantages, significantly enhancing PLS. We optimized secrecy rate through a two-stage approach, achieving joint power allocation and analog beamformer design with frameworks such as AO, FP, and BSUM. We also proposed a low-complexity beamsplit-aware beamfocusing method leveraging geometric information in near-field wideband propagation. Our results confirmed the superiority of these analog beamfocusing strategies over TTD-free approaches and highlighted the advantages in terms of hardware efficiency.

\appendices

\bibliographystyle{IEEEtran}
\bibliography{IEEEabrv,mybib}

\end{document}